# Non-Destructive Carotenoid Quantification in Leaves via Raman Spectroscopy: Optimizing Treatment for Linear Discriminant Analysis

Miri Park*[1], Annette Somborn[1], Dennis Schlehuber[1], Hyun Jeong Lim[2], and Volkmar Keuter[1]

[1]Fraunhofer Institute for Environmental, Safety and Energy Technologies UMSICHT, Oberhausen, Germany

[2]Department of Environmental Science and Engineering, Ewha Womans University, Seoul, Republic of Korea

\* Correspondence: miri.park@umsicht.fraunhofer.de

**This PDF file includes:**

    Main manuscript

    Figures

    References

NON-DESTRUCTIVE CAROTENOID QUANTIFICATION IN LEAVES VIA RAMAN SPECTRPSCPOPY

**ABSTRACT**

This study introduces a novel method for quantifying challenging carotenoids in leaf tissues, which typically produce less stable signals than fruits, grains, and roots, by applying Linear Discriminant Analysis (LDA) modeling to interpret Raman spectroscopy data. The model's performance was assessed across different spectral preprocessing techniques (smoothing, normalization, baseline correction) and through various subsets of relevant Raman shifts. To generate a broad range of carotenoid contents, genetically modified *Arabidopsis thaliana* mutants with controlled synthesis and more conventional *Spinacia oleracea* samples under dark and salt stress were utilized, allowing for the evaluation of model robustness and practical applicability. Transition scores for *Arabidopsis thaliana* reached 77.27–95.45% in all quantifications, while *Spinacia oleracea* showed 75–83.33% in 3- and 4-level modeling, demonstrating the LDA model's strong potential for effective application. Among the spectral preprocessing, smoothing had the greatest impact on model performance, enhancing results for *Arabidopsis thaliana* but showing better outcomes without smoothing for *Spinacia oleracea*. Overall, this study highlights the potential of LDA modeling combined with Raman spectroscopy as a robust and non-destructive tool for metabolite quantification in herbal plants, with promising applications in agricultural monitoring and quality control.

**SIGNIFICANCE**

The use of Raman spectroscopy for non-destructive, real-time evaluation of carotenoids content in agriculture has gained considerable attention. However, its application has been limited for crops consumed as leaves, which motivated us to develop a new approach to interpret the Raman signal. Through advanced Linear Discriminant Analysis modeling, we present a more accurate and practical approach for quantifying carotenoids content. This method holds high potential in that it not only enables carotenoids quantification but also offers a more general and versatile approach for broader use in various agricultural applications.

**INTRODUCTION**

Carotenoids, including lutein, zeaxanthin, lycopene, and α- and β-carotene, are natural lipid-soluble pigments synthesized in fruits and vegetables. Beyond their role in plant pigmentation, carotenoids serve as key biomarkers of plant health, with their levels fluctuating in response to nitrogen deficiency, abiotic stress, and pathogen infections, making them valuable indicators for monitoring plant well-being (1–3). Additionally, due to their potent antioxidant properties, which contribute to anti-inflammatory effects, immune system support, and anti-aging, carotenoids play a crucial role in human health and nutrition. As both plant health markers and beneficial dietary compounds, carotenoids continue to attract interest in agriculture and food sciences, particularly among farmers seeking efficient crop monitoring tools. However, achieving high carotenoid content in crop products is challenging for two primary reasons. First, the biosynthetic pathways of carotenoids in crops are highly dynamic and can fluctuate significantly throughout the cultivation cycle (4). Second, conventional methods for quantifying carotenoid content, which involves harvesting a portion of the crop, extracting carotenoids, and analyzing them using liquid chromatography or mass spectrometry, are both labor-intensive and time-consuming. These destructive methods also lack the ability to provide real-time monitoring of carotenoid content, limiting their utility for dynamic assessment during cultivation, highlighting the need for technology that can monitor changes in carotenoid content in real-time (5, 6).

Among advancements in non-destructive and real-time techniques, Raman spectroscopy for carotenoid quantification is considered promising because it can directly detect the molecular bond signals of specific target substances, even in small sample portions, and remains unaffected by the presence of water (7, 8). To date, Raman spectroscopy has advanced in studying the identification and quantification of carotenoid content in plants; however, research has primarily focused on fruits and roots with high accumulation (5, 9, 10). In studies involving leaves, most applications have utilized carotenoid signals to detect stress, ripeness, or disease rather than for direct quantification (1, 2, 11). Although there are a few cases of carotenoid quantification in plant leaves (12), its accuracy is relatively lower than in fruits, especially for specific carotenoid compounds rather than total carotenoids, as structurally similar carotenoids—β-carotene, lutein, zeaxanthin, and lycopene—exhibit nearly identical Raman signatures, with only slight frequency shifts related to polyene chain length (13). Additionally, Partial Least Squares Regression (PLSR), the preferred method for quantifying target compounds in most cases involving spectral data (14), models spectral data by identifying statistical correlations rather than directly resolving individual spectral features. While it is effective when the target compound accumulates at high concentrations, producing strong signals with minimal interference from other substances (15, 16), it struggles in complex systems like plant leaves, where carotenoid and chlorophyll Raman signals overlap, reducing quantification accuracy (12). To mitigate chlorophyll fluorescence, a 785 nm or near-infrared excitation wavelength was employed; however, this adaptation does not fully resolve interference issues, particularly in chlorophyll-rich samples (13). These limitations underscore the need for more targeted analytical approaches beyond conventional PLSR for leafy crops.

In this context, we introduce a novel quantification method using Linear Discriminant Analysis (LDA). LDA, originally developed by Ronald A. Fisher in 1936 for the taxonomic classification of Iris species (*Iris setosa*, *Iris versicolor*, and *Iris virginica*) based on their features (17), is a statistical technique that aims to find a linear combination of features that best separates predefined classes. This method continues to be widely used today for plant phenotyping (18) and assessing plant diversity (19) by applying it to various types of spectral data. Even though case studies assessing the LDA model for quantifying metabolites in plants are still untapped, it has potential in that LDA offers a distinct advantage by categorizing samples based on their inherent characteristics (20), a capability particularly valuable in agricultural settings where classifying target plants into predefined quality levels is often more practical than predicting exact component concentrations.

Here, we propose an application of LDA by leveraging its explicit modeling of inter-class (here, range of carotenoids contents) differences, a feature absent in conventional methods like Principal Component Analysis (PCA) or PLSR, to quantify specific carotenoids in agricultural crops leafy samples using Raman spectra, and analyze the factors affecting its performance to present optimal spectral preprocessing and analysis results. By focusing on class distinctions rather than precise quantification, this approach offers a unique perspective for rapidly and efficiently assessing crop quality. Using *Arabidopsis thaliana* mutants, we established the validity of the fundamental model and further evaluated its practical applicability to *Spinach oleracea* in real agricultural settings.

## RESULT

The carotenoid content for each line of *Arabidopsis thaliana* is shown in *SI Appendix*, **Table S1**. For *Spinach oleracea*, details of stress adaptation and changes in carotenoid content are provided in *SI Appendix*, **Table S2**. *SI Appendix*, **Table S3** presents the carotenoid ranges and mean values for each level divided by k-means clustering, as well as the number of samples used for modeling.

**Quantification of Carotenoids Content in Mutant *Arabidopsis thaliana***

*Lutein*

In *Arabidopsis thaliana*, where carotenoid synthesis was artificially modified using mutants, the best LDA models achieved a transition score exceeding 80% across all levels, with a particularly high score of 95.45% in the 3-level modeling scenario (**Fig. 1f-h**). Correlation analysis between spectral preprocessing methods and LDA model performance revealed that soothing and normalization improved model outcomes compared to the data without any treatments. Compared to the reference, all smoothing methods increased the transition score by approximately 8% (**Fig. 1a**), while normalization methods N01 and SNV enhanced the transition score by about 5% (**Fig. 1b**). Consistent with these findings, prediction and accuracy scores also showed corresponding improvements (*SI Appendix*, **Table 4**). Baseline correction using a third-order polynomial (B3) demonstrated relatively high LDA model performance (**Fig. 1c**). However, analysis of variance (ANOVA) testing indicated that the differences were not statistically significant (*SI Appendix*, **Table S4**). **Fig. 1e** illustrates the impact of spectral preprocessing combinations on LDA modeling performance, using the combination of SX, NormX, and B3 as references. Consistent with previous findings, the combinations of S1-S3, N01 and SNV, and B3 showed enhanced performance. Notably, the S3_N01_B3 combination improved the transition score by an average of 18.33% over the reference, with corresponding increases in prediction and accuracy scores (*SI Appendix*, **Table S5**). Subset selection for modeling significantly impacted model performance, emphasizing the critical role of choosing appropriate data subsets for robust LDA modeling. Defining subsets based on peak regions derived from lutein resulted in improved model performance across all cases, with transition score enhancements ranging from 4.71% to 19.28% compared to models built using the full Raman shift range (**Fig. 1d** and *SI Appendix*, **Table S4**). Particularly, Lut2 (1100–1240 cm$^{-1}$) and the broader regions Lut6 (980–1040 & 1100–1240cm$^{-1}$), 8 (1100–1240 & 1500–1550cm$^{-1}$), and 9 (980–1040 & 1100–1240 & 1500–1550 cm$^{-1}$), which include this range, showed high performance. In the 3 and 4-level modeling, Lut6 was included in the best-performing combinations, demonstrating its significant contribution to model prediction, transition scores, and accuracy (**Fig. 1f** and **g**).

*β-carotene*

The quantification results of β-carotene using the best LDA model showed relatively lower prediction, transition score, and accuracy compared to lutein (**Fig. 2f-h**). However, both 3-level and 4-level modeling achieved transition scores exceeding 85%. Even in the case of 5-level modeling, the confusion matrix in **Fig. 2h** demonstrates that a considerable number of samples were predicted within the actual range or a closely related range, highlighting the model's reasonable performance. Similar to lutein, the β-carotene models also showed improved performance when built using spectra that had undergone spectral smoothing and normalization processes. Specifically, smoothing increased the transition score by approximately 3%, while normalization contributed an additional 5% improvement (**Fig. 2a** and **b**, and *SI Appendix*, **Table S6**). Unlike lutein, baseline correction with a 5th-order polynomial (B5) yielded relatively higher transition scores for β-carotene. However, ANOVA analysis revealed that baseline correction did not result in statistically significant differences in prediction, transition score, or accuracy (**Fig. 2c** and *SI Appendix*, **Table S6**). Nevertheless, the combination of smoothing, normalization, and B5 baseline correction enhanced the transition score by 5.37% to 9.78% compared to the reference preprocessing method SX_NormX_B3 (52.07%) (**Fig. 2e** and *SI Appendix*, **Table S7**), demonstrating the potential benefit of more advanced preprocessing techniques. For subset selection, Bet8 (1040–1080, 1530–1600 cm$^{-1}$) demonstrated the best LDA performance (90.91% of transition score for 3-level modeling, 86.36% for 4-level, and 77.27% for 5-level modeling). Unlike lutein quantification, the performance of β-carotene quantification model improved when broader spectral ranges, such as the full reference range of 800–1750 cm$^{-1}$ or subsets like Bet4 (1040–1280cm$^{-1}$) and 5 (1040–1600cm$^{-1}$), which included not only β-carotene-derived peak areas but also additional regions between these peaks, were used in modeling (**Fig. 2d**). In contrast, models considering only the specific peak areas derived from β-carotene showed comparatively lower transition scores. Interestingly, for β-carotene, while normalization and subset selection enhanced model performance in terms of prediction and transition scores, they also showed a slight tendency to reduce accuracy (*SI Appendix*, **Table S6**).

**Quantification of Carotenoids Content in *Spinach oleracea***

To evaluate whether the LDA models developed and tested using mutants could be applied to actual crops, additional modeling was conducted on spinach for quantifying lutein and β-carotene. In lutein quantification, the LDA models demonstrated transition scores of 83.33% and 75% in the 3-level (**Fig. 3b**) and 4-level models (**Fig. 3c**), respectively. For β-carotene, the 3-level model achieved a transition score of 75% (**Fig. 3e**), while the 4-level model reached 81.25% (**Fig. 3f**).

The results revealed distinct differences in how spectral preprocessing influenced LDA model performance for each component. Notably, the effect of smoothing showed a clear divergence. In both lutein and β-carotene quantification, LDA models built using SX—without any smoothing processes—achieved significantly higher transition scores compared to models incorporating smoothing techniques (*SI Appendix*, **Table S8** and **S9**). Normalization also improved model performance; however, the effect was not statistically significant in the quantification of β-carotene (*SI Appendix*, **Fig. 1b** and **f**). Overall, the combination of SX and either N01 or SNV spectral preprocessing increased the prediction score by up to 8.61% for lutein and 5.94% for β-carotene, demonstrating the potential of tailored preprocessing strategies to enhance model accuracy (**Fig. 3a** and **d**, and *SI Appendix*, **S10** and **S11**). For lutein, similar to the results observed in *Arabidopsis thaliana*, robust models were generated in subsets Lut6 and Lut8-9, which included the 1100–1240 cm$^{-1}$ region, even in the 5-level model with 68.75% of transition score (***SI Appendix***, **Fig. S2**). In contrast, for β-carotene, modeling performance showed no significant increase in model performance across all subsets, indicating a consistent but less dynamic response to subset selection (***SI Appendix***, **Fig. S1d** and **h**).

## DISCUSSION

Traditional Raman spectroscopy methods for quantifying carotenoids in crop products, particularly leaves, presented prediction limitations and challenges. Within this context, we demonstrated the significant potential of LDA methods, which involves predicting the range of each target substance, to overcome these challenges. Additionally, we evaluated the effectiveness of the model by assessing how the spectrum processing methods and the subset selection approach, depending on the target crop and the type of carotenoids to be detected, influenced the model's performance.

Although direct comparison between the PLSR model, which predicts exact values rather than concentration ranges, and the validation performance of our proposed LDA model is challenging, our results clearly demonstrate the superior performance of the LDA model. When utilizing the same dataset presented in this study, the PLSR models developed for carotenoid quantification in both *Arabidopsis thaliana* and *Spinacia oleracea* exhibited significantly lower accuracy and validation values compared to our proposed model, with the sole exception being the model quantifying lutein in *Arabidopsis thaliana* mutant, which achieved acceptable accuracy (***SI Appendix***, **Fig. S3**). Notably, the PLSR model entirely failed to predict unknown samples for *Spinacia oleracea*, underscoring its limitations in broader application scenarios. These findings emphasize the high potential of LDA model for practical crop analysis, demonstrating robust performance in carotenoid quantification across diverse sample sets.

When examining the impact of spectral preprocessing on model performance, the effects of normalization and baseline correction appeared to be generally similar or exhibited no clear trend regardless of plant species or target metabolites. In contrast, the smoothing factor had opposite effects on the *Arabidopsis thaliana* and *Spinacia oleracea* sample groups. This discrepancy can be reasonably interpreted considering that smoothing reduces noise signals. As shown in ***SI Appendix***, **Fig. S4**, for *Arabidopsis thaliana*, which uses a mutant strain with distinctly different carotenoid content levels, the inherent discriminative features between levels were already well-recognized. Therefore, noise reduction through smoothing contributed to improved model performance. Conversely, in the case of *Spinacia oleracea*, where cultivation stress was applied to a conventional cultivar resulting in a normal distribution of concentrations, smoothing had an undesirable effect by also diminishing the underlying Raman features (21). It suggests that, in practical applications, avoiding smoothing may enhance model performance. Similarly, the impact of subset selection on the model can be interpreted along the same manner. In the mutant samples, the differences in Raman signals originating from each component were distinctly recognized at each level, likely resulting in a more pronounced improvement in the model's robustness through subset selection. This finding highlights the importance of tailoring preprocessing strategies to specific application scenarios, particularly when transitioning from controlled mutant studies to real-world crop analysis.

The proposed model achieved promising performance, as indicated by the transition score. However, in some cases, its performance in prediction showed room for improvement in certain cases, highlighting the potential for further applications of the LDA model. In this study, a basic approach was adopted to evaluate the feasibility of applying the LDA model. Therefore, there remain significant untapped ways to enhance the model's performance. Future research should explore promising directions, including following key directions. First, expanding the range of preprocessing techniques—such as multiplicative scatter correction (MSC), inverse MSC, wavelet transform (WT), or even harnessing deep learning for advanced noise reduction—could substantially elevate model performance (22–24). Additionally, the LDA model can be integrated with various applications (25), which also presents a potential avenue for further exploration. Ultimately, these advancements could not only enhance model performance but also unlock new possibilities for broader applications as a sensor in agriculture, enabling more precise monitoring.

This study presents a novel approach using spectral features in LDA model to quantify carotenoids content in plant products, being attempted for the first time and the method demonstrates considerable potential for practical applications in agriculture. In practical agricultural settings, particularly when assessing target metabolites content, it is generally more efficient to categorize samples into predefined concentration ranges rather than aiming for precise quantification. This approach not only aligns well with industry practices but also enhances the practical applicability of the proposed method, positioning the LDA model as a highly promising tool for real-time agricultural analysis. Moreover, a key advantage of this method is its broad versatility, as it is not limited to carotenoids quantification but can be effectively adapted to detect a wide range of target metabolites, expanding its utility across diverse agricultural contexts.

## METHODS

### Plant Samples

*Arabidopsis thaliana*

Seeds were obtained from the Nottingham Arabidopsis Stock Centre (NASC, Leicestershire, UK) under the following stock numbers: N1092, N505018, N616660, and N2104258. Stock N1092, corresponding to the standard wild-type (Wt, Col-0), was used as the control exhibiting the canonical carotenoid profile under our growth conditions. In contrast, the remaining stocks are T-DNA insertion lines reported to harbor disruptions in genes involved in carotenoid metabolism. Stock N505018, also known as Lutein-deficient 2 (*lut2*), fails to synthesize lutein due to the inactivation of the lycopene ε-cyclase gene (26). Stock N616660, designated Lutein-deficient 5 (*lut5*), leads to an accumulation of α-carotene, thereby reducing lutein synthesis. Moreover, some studies suggest that the increased α-carotene occupancy on protein–pigment complexes displaces β-carotene from its binding sites, ultimately decreasing β-carotene accumulation (27). Stock N20104258, identified as Carotenoid and Chloroplast Regulation 2 (*ccr2*), also affects carotenoid metabolism, leading to alterations in the accumulation of both lutein and β-carotene (28). After surface sterilization in ethanol, the seeds were sown in soil and cultivated in a chamber at 20°C for 6 weeks.

NON-DESTRUCTIVE CAROTENOID QUANTIFICATION IN LEAVES VIA RAMAN SPECTRPSCPOPY

*Spinach oleracea*

Spinach, obtained from the German company 'Sperli', was sown in soil and cultivated in a plant chamber under a 16:8-hour day-night cycle (PPFD: 160 µE·m$^{-2}$·s$^{-1}$), at 18°C and 70% relative humidity. The plants were watered every second day and harvested at 10 weeks of age, following germination. To achieve a broader range of carotenoid content, spinach was subjected to darkness and salt stress (300 mM of NaCl) for specified durations (1, 3, 7, and 10 days) prior to harvest, based on the case studies that demonstrated alterations in lutein and β-carotene synthesis under these conditions (29–31). A total of 89 individual plants were divided into 13 groups.

**Raman Spectroscopy Measurements**

All measurements were conducted using a 785 nm laser and ×5 magnification with Raman spectroscopy (inVia Qontor, Renishaw, Wotton-under-Edge, England, UK), controlled by Renishaw's WiRE software. The collected Raman shift range was 770–1850 cm$^{-1}$. For consistency in measurement, two leaves from each individual plant were selected for Raman measurements: the outermost (oldest) leaf and the innermost (youngest) leaf. Spectra were acquired from six distinct measuring points on each leaf, with a minimum spacing of 100 µm between points to mitigate potential laser exposure effects. Measurement areas were carefully selected to exclude the leaf lamina. A low-energy laser with a power of 150 mW was used, with an exposure time of 2.5 seconds per measurement to minimize fluorescence interference. Each spectrum was accumulated 10 times to enhance measurement accuracy. **Fig. 4a** presents a schematic of the measurement setup.

**High-Performance Liquid Chromatography (HPLC) Analysis**

High-performance liquid chromatography (HPLC) was adapted as the reference method. All aerial parts of the sample plants, except for the two discs used for Raman measurement, were extracted for HPLC analysis. The harvested samples were immediately frozen in liquid nitrogen and then freeze-dried overnight using a vacuum dryer (Martin Christ, Osterode am Harz, Germany) at -20°C. The dried samples were homogenized with 10% n-hexane (>99%, Merck) in ethanol (>99.8%, VWR) and sonicated for 90 minutes, with the temperature carefully maintained below 40°C during the process. The extraction was filtered using a cellulose filter with a retention range of 5–8 µm. HPLC analysis was carried out using an AZURA® HPLC 862 system (Knauer, Berlin, Germany), equipped with an Agilent Poroshell 120 EC-C18 column (4.6 × 150 mm, 4 µm). HPLC-grade Acetone (Sigma-Aldrich) and ultrapure water (Milli-Q, Merck, resistivity 18.2 MΩ·cm at 25°C) were employed as the mobile phase eluents. The carotenoids standard solutions were prepared with the same lutein (Thermo Scientific, CAS 127-40-2) and β-carotene (BLDpharm, CAS 7235-10-7) powder. A calibration curve was established within the concentration range of 0.8 ppm to 100 ppm. As an internal standard, *trans*-β-Apo-8′-carotenal (Sigma-Aldrich, CAS 1107-26-2) was used. Details of the HPLC methods are described in ***SI Appendix*, Text**.

**Data and Statistical Analysis**

*Spectral preprocessing*

In this study, to ensure the generalizability and simplicity of the modeling approach, commonly known and widely applied spectral processing methods were selected (23). These included smoothing to reduce noise, baseline correction to eliminate fluorescence effects— a common issue in biological sample measurements (32, 33).— and normalization to facilitate reliable cross-sample comparison. Four smoothing methods, including the 'no smoothing (SX)' option, three normalization techniques, including the 'no normalization (NormX)' approach, and three baseline correction methods were selected. The combined application of these methods resulted in 36 different processing strategies, whose effects on LDA modeling were thoroughly evaluated. Descriptions of each processing method and the polynomial equations used for baseline correction are presented in **Fig. 4b** and **Fig. 4c**, respectively. As the final step of spectral processing, outliers were removed from the 12 spectra measured for each individual plant using principal component analysis (PCA) with a confidence level of 99%.

*Linear discriminant analysis (LDA)*

LDA modeling was used to quantify the content of target carotenoids, with LDA performed on the preprocessed spectra as described above. For modeling, 28 samples were used for modeling and 11 for validation in *Arabidopsis thaliana*, while 65 samples were used for modeling and 24 for validation in *Spinach oleracea*, with the carotenoids content ranges used for modeling and validation presented in ***SI Appendix***, **Table S12** and ***SI Appendix***, **Fig. S4**. LDA models were built at three, four, and five levels for each plant species and types of carotenoids (lutein and β-carotene), respectively. To achieve effective and objective clustering of concentration ranges at each level, the k-means clustering algorithm was applied using actual content values calculated with the reference HPLC method. K-means clustering was performed, based on Hartigan and Wong's algorithm (34). LDA models were built for all 36 combinations of spectral preprocessing methods, with each model also developed for different subsets based on the type of carotenoids. The subsets, shown in **Fig. 4c**, were generated by selecting peak regions associated with lutein or β-carotene. All constructed models were evaluated using three key criteria, all expressed as percentages: accuracy (the proportion of correctly classified samples), prediction (the model's capability to precisely estimate component concentrations), and transition score (a measure of near-accurate predictions, assigning a partial score of 0.5 to samples predicted in the adjacent level to their actual content in the confusion matrix).

*Evaluation of the impact of spectral preprocessing and subset on modeling performance*

The impact of spectral preprocessing and subset selection on the transition scores of LDA models was evaluated according to plant species and carotenoid types. The transition score was chosen as a key performance metric due to its ability to capture near-accurate predictions, offering a more nuanced assessment of model performance beyond simple accuracy or prediction. Linear modeling techniques (35) were applied to analyze the relationships between transition scores and subsets, as well as the effects of different preprocessing methods and their combinations. In the linear models, SX for smoothing, NormX for normalization, B3 for baseline correction, and the SX_NormX_B3 combination for preprocessing were selected as references. The full range of Raman shifts (800–1750 cm$^{-1}$) was designated as the reference for subset evaluations.

All analyses were conducted using R statistical software (R Core Team, version 4.3.2, Vienna, Austria).

**AUTHOR CONTRIBUTIONS**

MP structured the concept of this study, generated all data, and drafted the manuscript. AS, DS, HJL and VK contributed to the critical revision of the manuscript.

**FUNDING**

This study was partially supported by the Federal Ministry of Education and Research (BMBF) through the program Agricultural Systems of the Future_2 in the framework of the "National Research Strategy BioEconomy 2030" under grant No. 031B1528A "SUSKULT – Development of a Sustainable Cultivation System of Resilient Metropolitan Regions" and through the program "Model Region Bioeconomy in the Rhenish mining area" under grant No. 031B1137BX "Model region, Phase 1, BioRevierPLUS: InnoLA, TP2-circular PhytoREVIER"

**CONFLICT OF INTEREST**

The authors declare that they have no conflict of interest.

**FIGURE CAPTIONS**

**Fig. 1. Effects of spectral preprocessing and subset selection on LDA modeling performance for predicting lutein content in *Arabidopsis thaliana*** (**a–e**) Bar charts illustrating the impact of spectral preprocessing and subset selection on LDA modeling using transition scores derived from linear regression. (**a**), (**b**), (**c**), and (**d**) represent the effects of smoothing, normalization, baseline correction, and subset selection, respectively, with SX, NormX, B3, and 800–1750 cm$^{-1}$ Raman shift region as their corresponding references. (**e**) Displays the combined effect of smoothing, normalization, and baseline correction, using SX_NormX_B3 as the reference, with a standard error of 3.37. Statistical significance levels are indicated on each bar: *** ($p < 0.001$), ** ($p < 0.01$), * ($p < 0.05$), and . ($p < 0.1$). (**f–h**) LDA modeling results for predicting β-carotene content at different levels. Each panel contains the LDA plot, prediction values, transition scores, accuracy values, and the confusion matrix, which compares actual and predicted levels. (**f**) Results for 3-level modeling, (**g**) results for 4-level modeling, and (**h**) results for 5-level modeling.

**Fig. 2. Effects of spectral preprocessing and subset selection on LDA modeling performance for predicting β-carotene content in Arabidopsis thaliana** (**a–e**) Bar charts illustrating the impact of spectral preprocessing and subset selection on LDA modeling using transition scores derived from linear regression. (**a**), (**b**), (**c**), and (**d**) represent the effects of smoothing, normalization, baseline correction, and subset selection, respectively, with SX, NormX, B3, and 800–1750 cm$^{-1}$ Raman shift region as their corresponding references. (**e**) Displays the combined effect of smoothing, normalization, and baseline correction, using SX_NormX_B3 as the reference, with a standard error of 3.18. Statistical significance levels are indicated on each bar: *** ($p < 0.001$), ** ($p < 0.01$), * ($p < 0.05$), and . ($p < 0.1$). (**f–h**) LDA modeling results for predicting β-carotene content at different levels. Each panel contains the LDA plot, prediction values, transition scores, accuracy values, and the confusion matrix, which compares actual and predicted levels. (**f**) Results for 3-level modeling, (**g**) results for 4-level modeling, and (**h**) results for 5-level modeling.

**Fig. 3. Effects of spectral preprocessing on LDA modeling performance for predicting lutein and β-carotene content in *Spinacia oleracea*** (**a**) and (**d**) Bar charts showing the impact of combined spectral preprocessing (smoothing, normalization, and baseline correction) on LDA modeling using transition scores from linear regression. SX_NormX_B3 was used as the reference, with standard errors of 2.64 and 2.52, respectively. (**a**) Represents lutein, (**d**) β-carotene. Statistical significance: *** ($p < 0.001$), ** ($p < 0.01$), * ($p < 0.05$), . ($p < 0.1$). (**b–c**) and (**e–f**) LDA modeling results for lutein and β-carotene content. Each panel includes the LDA plot, prediction values, transition scores, accuracy, and confusion matrix. (**b**) and (**e**) show 3-level modeling, and (**c**) and (**f**) show 4-level modeling with 3D LDA plots to efficiently visualize distributions.

**Fig. 4. Schematic overview of this study** (**a**) Schematic illustration of the Raman measurement setup used in this study. CCD (Charge-Coupled Device) detects scattered light, allowing spectral acquisition of plant leaves (**b**) Different types of spectral preprocessing methods applied before LDA modeling, including smoothing, normalization, and baseline correction. (**c**) Equations for baseline correction (**d**) Raman spectra of *Arabidopsis thaliana* wildtype (Wt), lutein, and β-carotene standards. The highlighted spectral regions—green for lutein and yellow for β-carotene—indicate the selected Raman shift ranges corresponding to vibrational modes characteristic of each carotenoid. These regions were specifically used to enhance LDA modeling performance by focusing on chemically relevant spectral features.

NON-DESTRUCTIVE CAROTENOID QUANTIFICATION IN LEAVES VIA RAMAN SPECTRPSCPOPY

**Fig. 1**

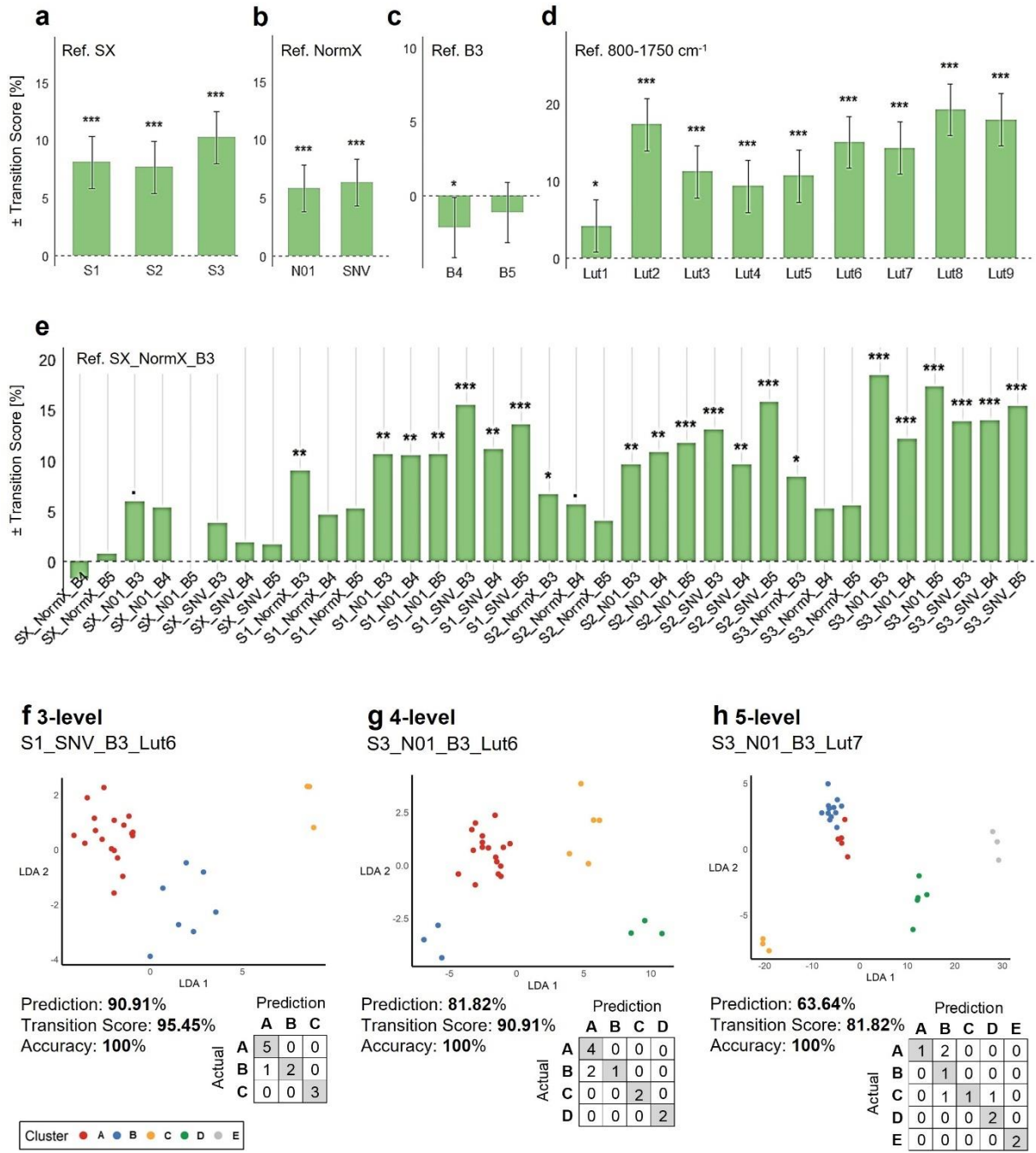

**Fig. 2**

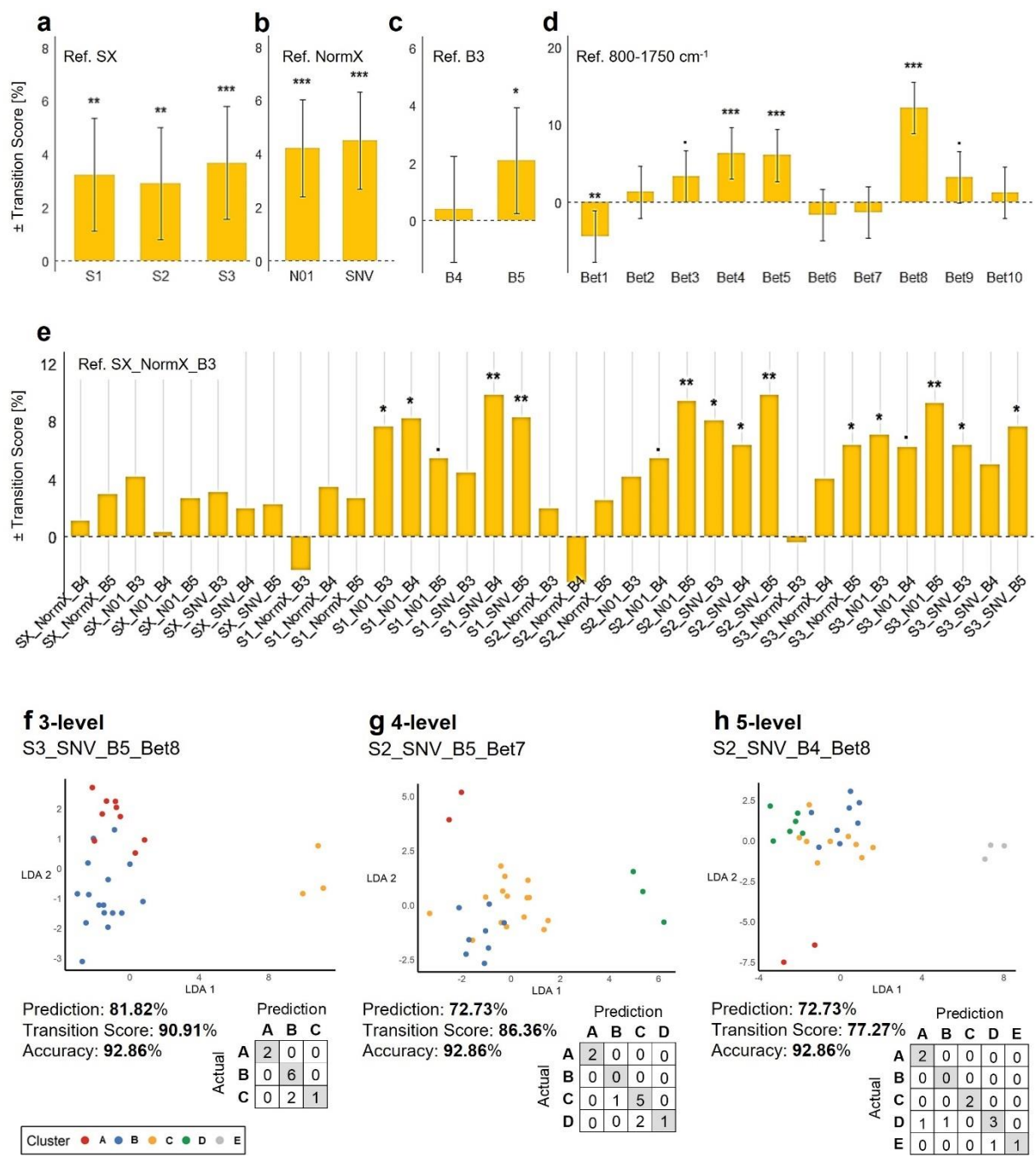

**Fig. 3**

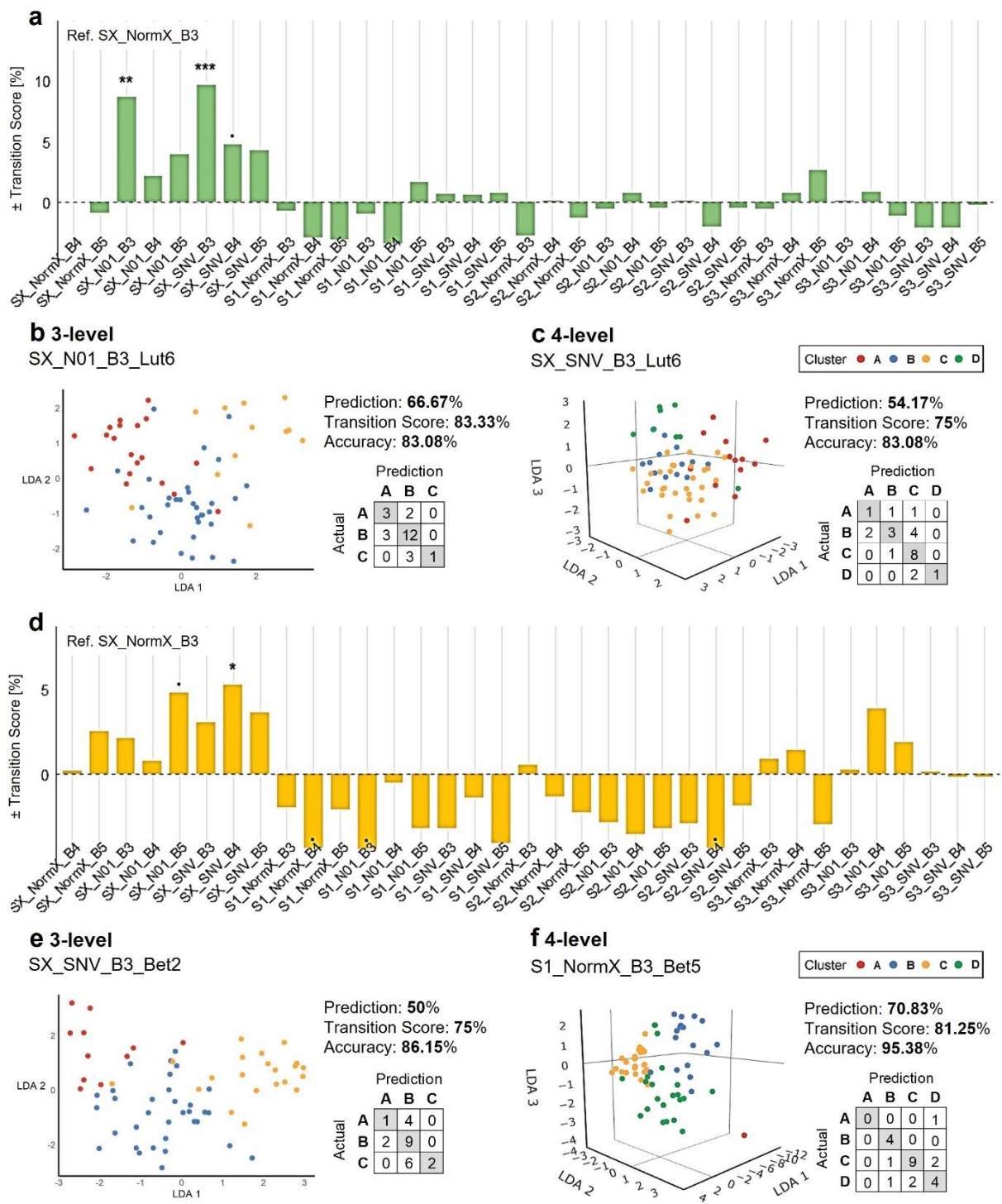

**Fig. 4**

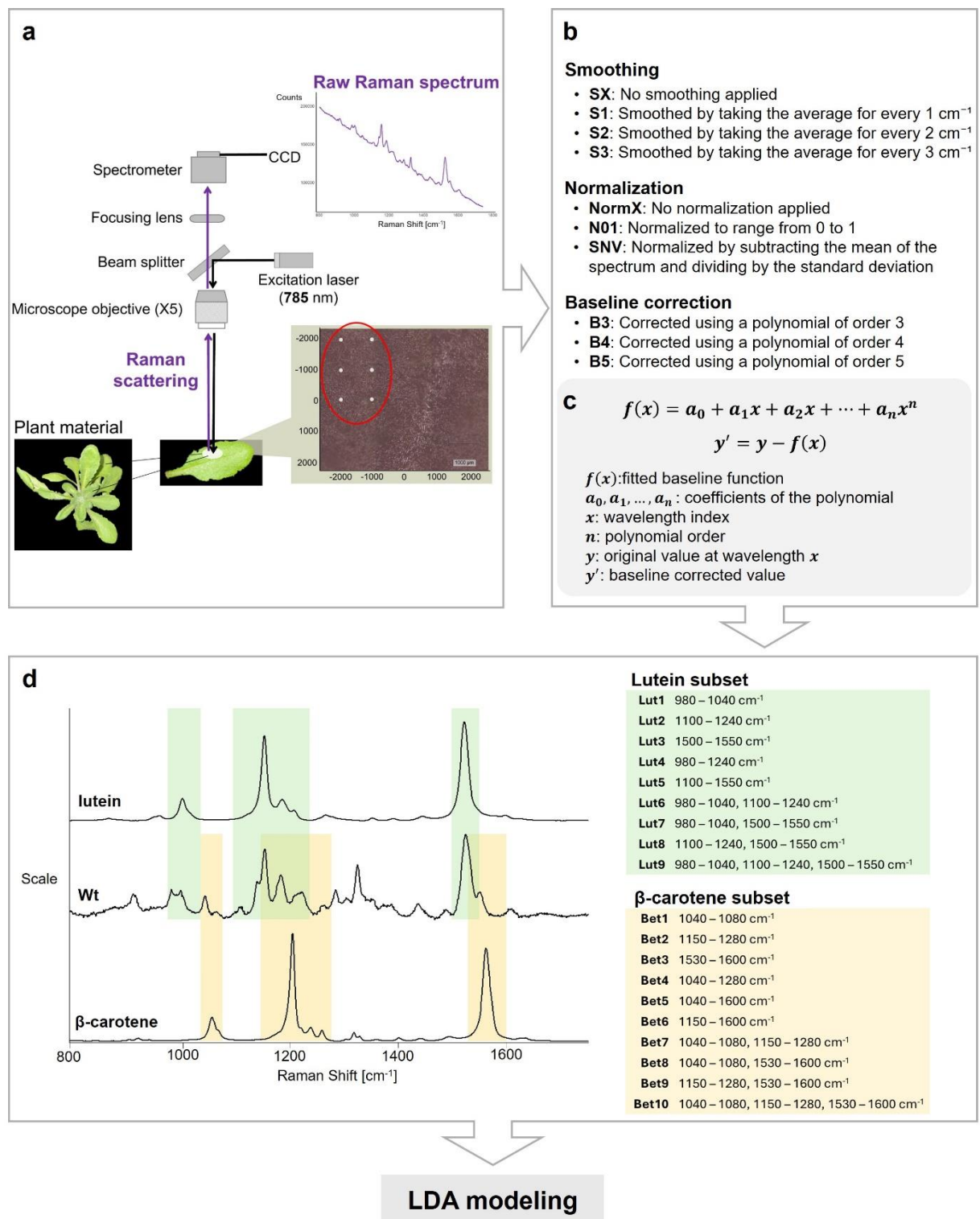

# REFERENCES

1. Gupta S et al. (2020) Portable Raman leaf-clip sensor for rapid detection of plant stress. Scientific Reports 10:20206.

2. Altangerel N et al. (2017) In vivo diagnostics of early abiotic plant stress response via Raman spectroscopy. Proceedings of the National Academy of Sciences 114:3393–3396.

3. Yeturu S et al. (2016) Handheld Raman spectroscopy for the early detection of plant diseases: Abutilon mosaic virus infecting Abutilon sp. Analytical Methods 8:3450–3457.

4. Brychkova G et al. (2023) Regulation of Carotenoid Biosynthesis and Degradation in Lettuce (Lactuca sativa L.) from Seedlings to Harvest. International journal of molecular sciences 24.

5. Park M, Somborn A, Schlehuber D, Keuter V, Deerberg G (2023) Raman spectroscopy in crop quality assessment: focusing on sensing secondary metabolites: a review. Hortic Res 10:uhad074.

6. Fu X, He X, Xu H, Ying Y (2016) Nondestructive and Rapid Assessment of Intact Tomato Freshness and Lycopene Content Based on a Miniaturized Raman Spectroscopic System and Colorimetry. Food Analytical Methods 9:2501–2508.

7. Schulz H, Baranska M (2007) Identification and quantification of valuable plant substances by IR and Raman spectroscopy. Vibrational Spectroscopy 43:13–25.

8. Kaiqiang Wang, Zonglun Li, Jinjie Li, Hong Lin (2021) Raman spectroscopic techniques for nondestructive analysis of agri-foods: A state-of-the-art review. Trends in Food Science & Technology 118:490–504.

9. Hara R, Ishigaki M, Kitahama Y, Ozaki Y, Genkawa T (2018) Excitation wavelength selection for quantitative analysis of carotenoids in tomatoes using Raman spectroscopy. Food Chemistry 258:308–313.

10. Badgujar PM, Wang Y-C, Cheng C-L (2021) A light-mediated study of carotenoids in carrots using resonance Raman spectroscopy. J Raman Spectrosc 52:2609–2620.

11. Jianwei Qin, Kuanglin Chao, Moon Sung Kim (2012) Nondestructive evaluation of internal maturity of tomatoes using spatially offset Raman spectroscopy. Postharvest Biology and Technology 71:21–31.

12. Zeng J et al. (2021) Quantitative visualization of photosynthetic pigments in tea leaves based on Raman spectroscopy and calibration model transfer. Plant Methods 17:4.

13. Payne TD et al. (2024) Identification and quantification of pigments in plant leaves using thin layer chromatography-Raman spectroscopy (TLC-Raman). Analytical Methods 16:2449–2455.

14. Chang C-W, Laird DA, Mausbach MJ, Hurburgh CR (2001) Near-Infrared Reflectance Spectroscopy–Principal Components Regression Analyses of Soil Properties. Soil Sci. Soc. Am. J. 65:480–490.

15. Baranska M, Schütze W, Schulz H (2006) Determination of lycopene and beta-carotene content in tomato fruits and related products: Comparison of FT-Raman, ATR-IR, and NIR spectroscopy. Analytical Chemistry 78:8456–8461.

16. Kim SW et al. (2015) Monthly metabolic changes and PLS prediction of carotenoid content of citrus fruit by combined Fourier transform infrared spectroscopy and quantitative HPLC analysis. Plant Biotechnology Reports 9:247–258.

17. Fisher RA (1936) THE USE OF MULTIPLE MEASUREMENTS IN TAXONOMIC PROBLEMS. Annals of Eugenics 7:179–188.

18. Wahabzada M et al. (2016) Plant Phenotyping using Probabilistic Topic Models: Uncovering the Hyperspectral Language of Plants. Scientific Reports 6:22482.

19. Jain E, Rose M, Jayapal PK, Singh GP, Ram RJ (2024) Harnessing Raman spectroscopy for the analysis of plant diversity. Scientific Reports 14:12692.

20. Locatelli I et al. (2025) Progress in quality assessment of Italian saffron. Scientific Reports 15:2175.

21. Barton SJ, Ward TE, Hennelly BM (2018) Algorithm for optimal denoising of Raman spectra. Analytical Methods 10:3759–3769.

22. Li Y et al. (2022) Spectral Pre-Processing and Multivariate Calibration Methods for the Prediction of Wood Density in Chinese White Poplar by Visible and Near Infrared Spectroscopy. Forests 13.

23. Rinnan Å, van Berg F den, Engelsen SB (2009) Review of the most common pre-processing techniques for near-infrared spectra. TrAC Trends in Analytical Chemistry 28:1201–1222.

24. Helin R, Indahl UG, Tomic O, Liland KH (2022) On the possible benefits of deep learning for spectral preprocessing. Journal of Chemometrics 36:e3374.

25. Zhao S, Zhang B, Yang J, Zhou J, Xu Y (2024) Linear discriminant analysis. Nature Reviews Methods Primers 4:70.

26. Popova AV, Stefanov M, Ivanov AG, Velitchkova M (2022) The Role of Alternative Electron Pathways for Effectiveness of Photosynthetic Performance of Arabidopsis thaliana, Wt and Lut2, under Low Temperature and High Light Intensity. Plants (Basel, Switzerland) 11.

27. Flores Hector E., Galston Arthur W. (1982) Polyamines and Plant Stress: Activation of Putrescine Biosynthesis by Osmotic Shock. Science 217:1259–1261.

28. Nisar N, Li L, Lu S, Khin NC, Pogson BJ (2015) Carotenoid Metabolism in Plants. Molecular plant 8:68–82.

29. Kim B-M, Lee H-J, Song YH, Kim H-J (2021) Effect of salt stress on the growth, mineral contents, and metabolite profiles of spinach. Journal of the science of food and agriculture 101:3787–3794.

30. Xu C, Mou B (2016) Responses of Spinach to Salinity and Nutrient Deficiency in Growth, Physiology, and Nutritional Value. Journal of the American Society for Horticultural Science J. Amer. Soc. Hort. Sci. 141:12–21.

31. Kim C-K, Eom S-H (2025) Light Controls in the Regulation of Carotenoid Biosynthesis in Leafy Vegetables: A Review. Horticulturae 11.

32. Keisaku Hamada et al. (2008) Raman microscopy for dynamic molecular imaging of living cells. Journal of biomedical optics 13:1–4.

33. Yakubovskaya E, Zaliznyak T, Martínez Martínez J, Taylor GT (2019) Tear Down the Fluorescent Curtain: A New Fluorescence Suppression Method for Raman Microspectroscopic Analyses. Scientific Reports 9:15785.

34. Hartigan JA, Wong MA (1979) Algorithm AS 136: A K-Means Clustering Algorithm. Journal of the Royal Statistical Society. Series C (Applied Statistics) 28:100–108.

35. Chamber JM, Hastie TJ (1992) Statistical Models in S (Wadsworth & Brooks/Cole, New York).